\documentclass[twocolumn]{revtex4-1}
\pdfoutput=1
\usepackage{amsmath}
\usepackage{amsfonts}
\usepackage{amssymb}
\usepackage{braket}
\usepackage{siunitx}
\usepackage{graphicx}
\usepackage{tikz}
\usepackage{calc}
\usepackage{float}
\usepackage{color}
\usepackage{mathtools}

\begin{document}

\title{Observing light-induced Floquet band gaps in the longitudinal conductivity of graphene}
\author{Lukas Broers$^{1,2}$ and Ludwig Mathey$^{1,2,3}$}
\affiliation{
$^{1}$Center for Optical Quantum Technologies, University of Hamburg, 22761 Hamburg, Germany\\
$^{2}$Institute for Laser Physics, University of Hamburg, 22761 Hamburg, Germany\\
$^{3}$The Hamburg Center for Ultrafast Imaging, Luruper Chaussee 149, 22761 Hamburg, Germany}
\begin{abstract}
We propose optical longitudinal conductivity as a realistic observable to detect light-induced Floquet band gaps in graphene. 
These gaps manifest as resonant features in the conductivity, when resolved with respect to the probing frequency and the driving field strength. 
We demonstrate these features via a dissipative master equation approach which gives access to a frequency- and momentum-resolved electron distribution. 
This distribution follows the light-induced Floquet-Bloch bands, resulting in a natural interpretation as occupations of these bands. 
Furthermore, we show that there are population inversions of the Floquet-Bloch bands at the band gaps for sufficiently strong driving field strengths. 
This strongly reduces the conductivity at the corresponding frequencies. 
Therefore our proposal puts forth not only an unambiguous demonstration of light-induced Floquet-Bloch bands, which advances the field of Floquet engineering in solids, 
but also points out the control of transport properties via light, that derives from the electron distribution on these bands. 
\end{abstract}
\date{\today}
\maketitle
Controlling solids with light constitutes a modern approach to induce novel functionalities. 
A specific framework within this broader effort is Floquet engineering. 
Floquet engineering refers to inducing dynamics that are captured by an effective Floquet Hamiltonian in a system by periodic driving. 
For a non- or weakly interacting system this approach describes effective single-particle states that form a natural basis for the driven system. 
These states are the Floquet-Bloch bands of the electrons, in analogy to the Bloch bands of the equilibrium system. 
These Floquet-Bloch bands can have qualitatively distinct features from the Bloch bands of the non-driven system~\cite{Hsieh,Sota,Schliemann,Seradjeh,Fiete}. 
A striking example are Floquet topological insulators~\cite{Demler,Galitski,Rudner}, for which applications in quantum computing and spintornics have been discussed in~\cite{Sameti,Nayak,Utic}. 
A specific realization is monolayer graphene illuminated with circularly polarized light, for which the band structure approaches the Haldane model in the high-frequency limit~\cite{Haldane, Aoki}. 
However, while the ground state of the equilibrium Haldane model forms indeed a topological insulator, which manifests in a quantized Hall conductance, the Hall conductance of optically driven graphene is not topologically quantized, but of geometric-dissipative origin, see~\cite{Nuske,McIver}. 
This observation is part of the larger challenge of an unambiguous detection of the Floquet-Bloch bands in a solid. 
We note that the existence of these bands and their Berry curvature in a periodically driven hexagonal lattice have been demonstrated in ultracold atom experiments~\cite{Weitenberg}. 
While signatures of Floquet-Bloch bands have been seen \cite{Gedik}, a smoking-gun in the transport measurements of solids is lacking. 

\begin{figure}[h!]
\centering 
\includegraphics[width=1.0\linewidth]{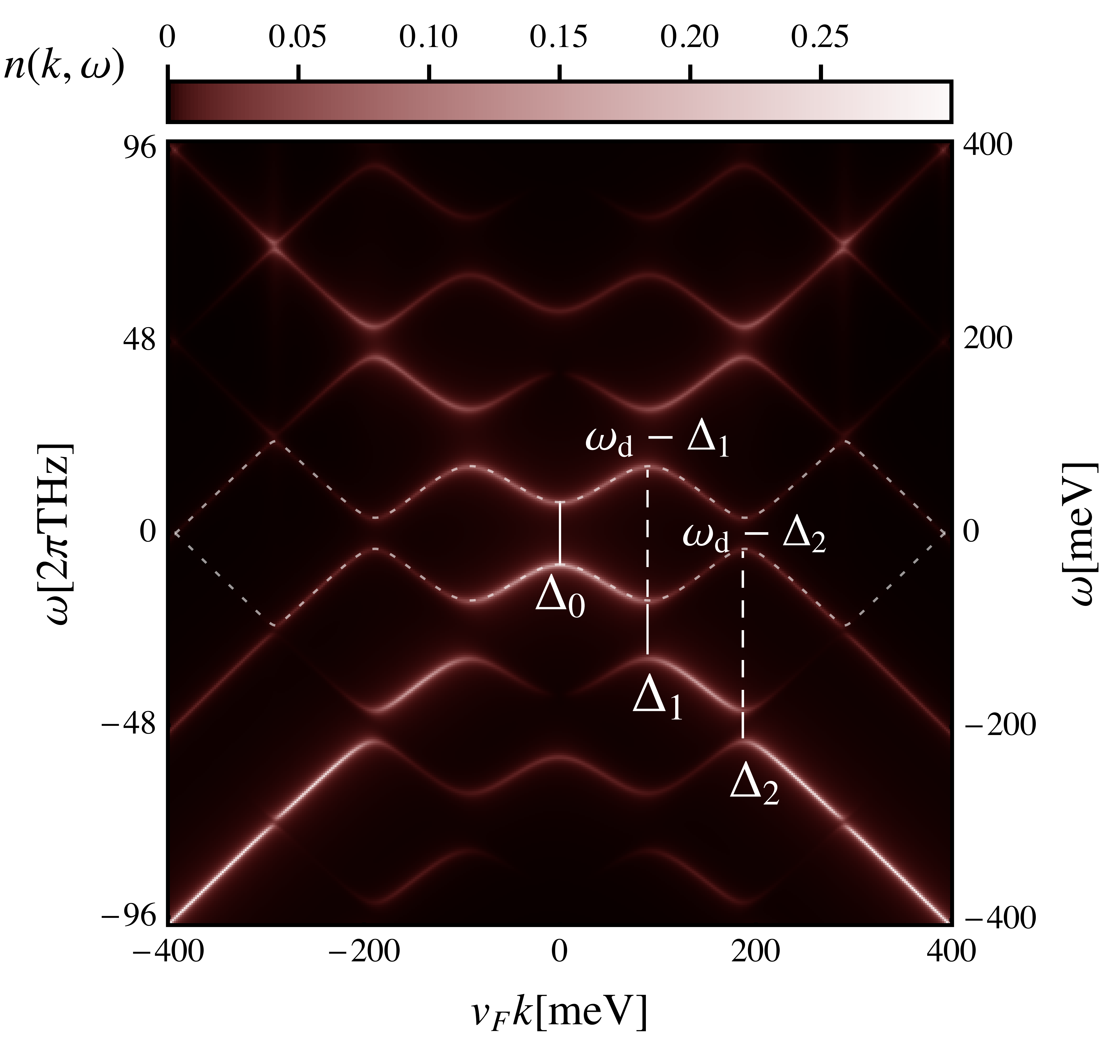}
\caption{The electron distribution $n({\bf k},\omega)$ of graphene driven with circularly polarized light at $\omega_\mathrm{d}=2\pi\times48\si{\tera\hertz}\approx 200\si{\milli\electronvolt}$ and $E_\mathrm{d}=26\si{\mega\volt\per\meter}$. 
The distribution $n({\bf k}, \omega)$ depends only on $k=|{\bf k}|$.
This quantity displays the steady state occupation of the Floquet-Bloch band structure. 
The one-photon resonance gap $\Delta_1$ at $k = \omega_\mathrm{d} /(2 v_F)$, the two-photon gap $\Delta_0$ at the Dirac point, and the two-photon gap at $k = \omega_\mathrm{d}/v_F$ are highlighted for clarity. 
Additionally, the complementary gaps $\omega_\mathrm{d} - \Delta_1$ and $\omega_\mathrm{d} - \Delta_2$ are indicated. The dotted lines indicate the Floquet energies of the first Floquet zone.}
\label{fig1}
\end{figure}

In this paper, we propose to detect light-induced Floquet band gaps in graphene via the optical longitudinal transport. 
We determine the optical conductivity as a function of the probing frequency and the driving field strength which displays resonant features.
We present an interpretation of these features in terms of the Floquet-Bloch band dispersion and the effective occupation of these states.
These occupations are determined by the dissipation and the driving field, which balance out to form the steady state.
We include the dissipation processes in our master equation approach that we use to describe the system.
With this we attribute the observable resonant features in the optical conductivity to two transition processes.
One occurs between bands inside the same Floquet zone and the other between adjacent bands of neighbouring Floquet zones.
These processes compete in their impact on the optical conductivity, which can result in vanishing and even negative optical conductivity for specific frequencies and driving field strengths.
In general we show that the conductivity depends on the relative occupation of the Floquet bands.
We also point out that the relative occupation is in qualitative agreement with a comoving band velocity, to be defined below.
In particular, we show that there are regimes of driving field strengths that show an effective inversion of Floquet band populations.
These are in the regimes in which negative optical conductivity is achieved.
Therefore, as a second point besides the demonstration of Floquet-Bloch bands in solids, our proposal shows non-trivial control of the transport properties of solids, induced by light.

We consider a circularly polarized laser with frequency $\omega_\mathrm{d}=2\pi\times 48\si{\tera\hertz}\approx 200\si{\milli\electronvolt}$ and variable field strength $E_\mathrm{d}$, which illuminates a graphene layer from perpendicular direction. 
The electromagnetic forces drive the electrons into a steady state.
We propose to measure the longitudinal AC conductivity of this steady state in the optical frequency domain.
The conductivity displays frequency regimes in which its magnitude is increased compared to the non-driven graphene layer, and regimes in which it is decreased. 
These frequency regimes derive from resonances between the Floquet states, which in turn depend on the driving field strength.
As a result, these frequency regimes can be tuned to overlap, resulting in a partial cancellation. 
In particular, the band gap $\Delta_0$ at the Dirac point can be overshadowed, in general, by other features. 
However, we point out a regime in which it can be identified unambiguously. 

\begin{figure}
\centering 
\includegraphics[width=1.0\linewidth]{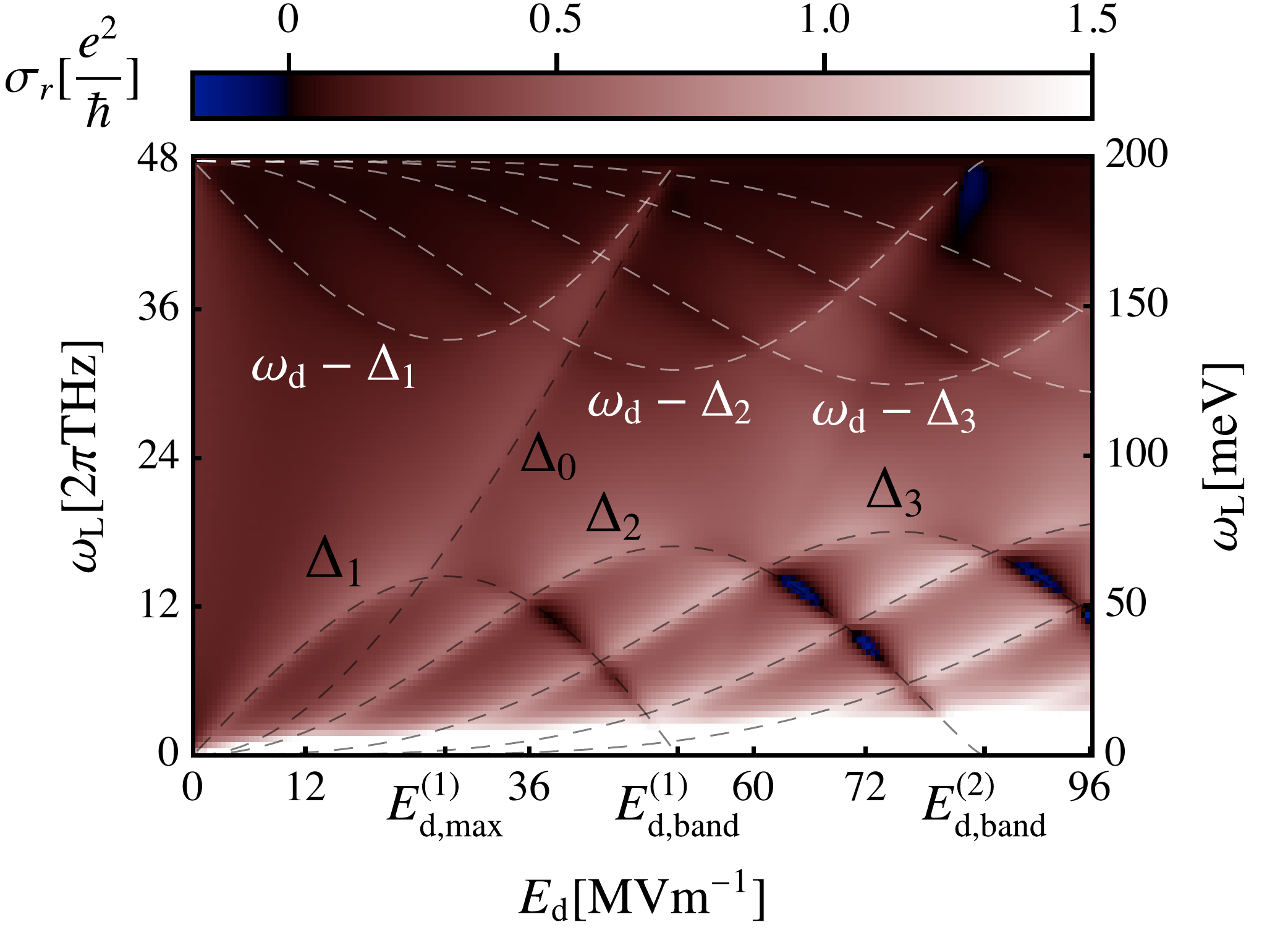}
\caption{The optical conductivity of graphene driven at $\omega_\mathrm{d}=2\pi\times 48\si{\tera\hertz}\approx 200\si{\milli\electronvolt}$ as a function of the driving field strength $E_\mathrm{d}$. 
The dashed lines show the various band gaps $\Delta_m$ as given by Floquet theory.
The gap $\Delta_0$ becomes clearly visible above values of $\omega_\mathrm{L}\approx 2\pi\times 14 \si{\tera\hertz} \approx 60 \si{\milli\electronvolt}$ and $E_\mathrm{d}\approx 28 \si{\mega\volt\per\meter}$. 
We also see the complementary resonant features at $\omega_\mathrm{d}-\Delta_{m}$, with $m>0$. }
\label{fig2}
\end{figure}

The Hamiltonian of light-driven graphene, close to the Dirac point is given by 
\begin{align}
H = \sum_k c_{\bf k}^\dagger \mathbf{h}({\bf k}) c_{\bf k},
\end{align}
where $c_{\bf k}= (c_{{\bf k},A}, c_{{\bf k},B})^\mathrm{T} $ and $c_{{\bf k},i}$ are the fermionic annihilation operators of an electron with momentum ${\bf k}$ and sublattice index $i=A,B$.
The Hamiltonian of a single momentum ${\bf k}$ is 
\begin{align}
\mathbf{h}({\bf k}) &= 
\hbar v_F (q_x \sigma_x + q_y \sigma_y)
\label{hamiltonian_graphene}
\end{align}
with
\begin{align}
q_x &=
k_x + \frac{E_\mathrm{d}}{\omega_\mathrm{d}}\sin(\omega_\mathrm{d} t) 
    - \frac{E_\mathrm{L}}{\omega_\mathrm{L}} \cos(\omega_\mathrm{L} t) \\
q_y &=
k_y+ \frac{E_\mathrm{d}}{\omega_\mathrm{d}}\cos(\omega_\mathrm{d} t) 
\end{align}
where $v_F\approx 10^6\si{\meter\per\second}$ is the Fermi velocity.
$k_i$ are the momentum components and $\sigma_i$ are the Pauli matrices. 
$E_\mathrm{d}$ and $\omega_\mathrm{d}$ are field strength and frequency of the driving laser. 
$E_\mathrm{L}$ and $\omega_\mathrm{L}$ are the same quantities for the longitudinal probing field.

We simulate the dynamics via a master equation approach, expanding on Ref.~\cite{Nuske}. The density matrix of the system factorizes in momentum space, as $\rho = \prod_{\bf k} \rho_{\bf k}$. Each $\rho_{\bf k}$ matrix operates on a four dimensional Hilbert space, given by the states $\ket{0}$, $c_{{\bf k},A}^\dagger\ket{0}$, $c_{{\bf k},B}^\dagger\ket{0}$, $c_{{\bf k},B}^\dagger c_{{\bf k},A}^\dagger\ket{0}$. 
We include doubly and unoccupied states to determine two-time correlation functions, and thereby frequency-resolved quantities. 

In addition to the unitary time evolution induced by the Hamiltonian in Eq.~\ref{hamiltonian_graphene}, we include dissipation via Lindblad operators defined in the instantaneous eigenbasis of the driven system, to describe the dissipative environment due to degrees of freedom not included in the Hamiltonian. 
We include a dephasing term $\gamma_z$, a decay term $\gamma_-$ and a term with decay rate $\gamma_\mathrm{bg}$ that models particle exchange of the graphene layer to a supporting substrate backgate.
This model provides a realistic discription of the non-equilibrium electron dynamics, see Ref.~\cite{Nuske}.

We choose the coefficients $\gamma_z=1\si{\tera\hertz}$, $\gamma_-=2.25\si{\tera\hertz}$ and $\gamma_\mathrm{bg}=2.5\si{\tera\hertz}$.
This sets the scale for the broadening of the effective bands in the single-particle correlation function as well as the optical conductivity. 
These values are a factor of $10$ smaller than those estimated for the experimental setup of Ref.~\cite{McIver} by Ref.~\cite{Nuske}.
Our predictions apply to high-mobility samples, e.g. BN-encapsulated graphene. 
For larger values, such as those that are realized in Ref.~\cite{McIver}, resolving the gap features that we describe in the following, would require larger driving frequencies and stronger driving. Throughout this work we use the temperature $T=80\mathrm{K}$ which is the same as the setup of Ref.~\cite{McIver}.

As a first observable we display the momentum- and energy-resolved electron distribution inspired by Ref.~\cite{Pruschke}
\begin{equation}
n({\bf k},\omega) = \int_{\tau_1}^{\tau_2} \int_{\tau_1}^{\tau_2} 
\mathcal{G}({\bf k},t_2,t_1) \frac{e^{i\omega (t_2-t_1)}}{(\tau_2-\tau_1)^2}
 \mathrm{d}t_2 \mathrm{d}t_1
\end{equation} 
with
\begin{equation}
\mathcal{G}({\bf k},t_2,t_1) = \braket{ c_{{\bf k},A}^\dagger(t_2) c_{{\bf k},A}(t_1)}+\braket{ c_{{\bf k},B}^\dagger(t_2) c_{{\bf k},B}(t_1)}.
\end{equation}
We use the time interval $[\tau_1, \tau_2]$ as the probing interval.
We choose $\tau_1$ such that the system has reached its steady state. 
$\tau_2-\tau_1$ is a sufficiently long probing time of the order of hundreds of driving periods $2\pi/\omega_\mathrm{d}$ that is also commensurate with the probing period $2\pi/\omega_\mathrm{L}$.
We note that this quantity provides a prediction for trARPES measurements~\cite{Pruschke}.
In Fig.~\ref{fig1} we show $n(k=|{\bf k}|,\omega)$ for the driving field strength $E_\mathrm{d}=26\si{\mega\electronvolt\per\meter}$. 
We note that a similar result was presented in Ref.~\cite{Nuske}.
The electron distribution of the steady state is consistent with the effective band structure predicted by Floquet theory and identifies the non-equilibrium electron occupation of these Floquet-Bloch bands.

We label the band gaps as $\Delta_m$, based on their location $m\omega_\mathrm{d}/(2 v_F)$ in momentum space for small driving field strength $E_\mathrm{d}\rightarrow 0 $, as shown in Fig.~\ref{fig1}.
Due to the periodicity in frequency space of the Floquet spectrum, there is a complementary gap $\omega_\mathrm{d}-\Delta_m$ for any given band gap $\Delta_{m}$, with $m>0$. 
These complementary gaps are also visible in the optical conductivity of the system. 
They reduce the conductivity at the corresponding frequency, rather than enhance it.
The gap $\Delta_0$ at the Dirac point does not exhibit this behavior, as discussed later.

\begin{figure}
\centering 
\includegraphics[width=1.0\linewidth]{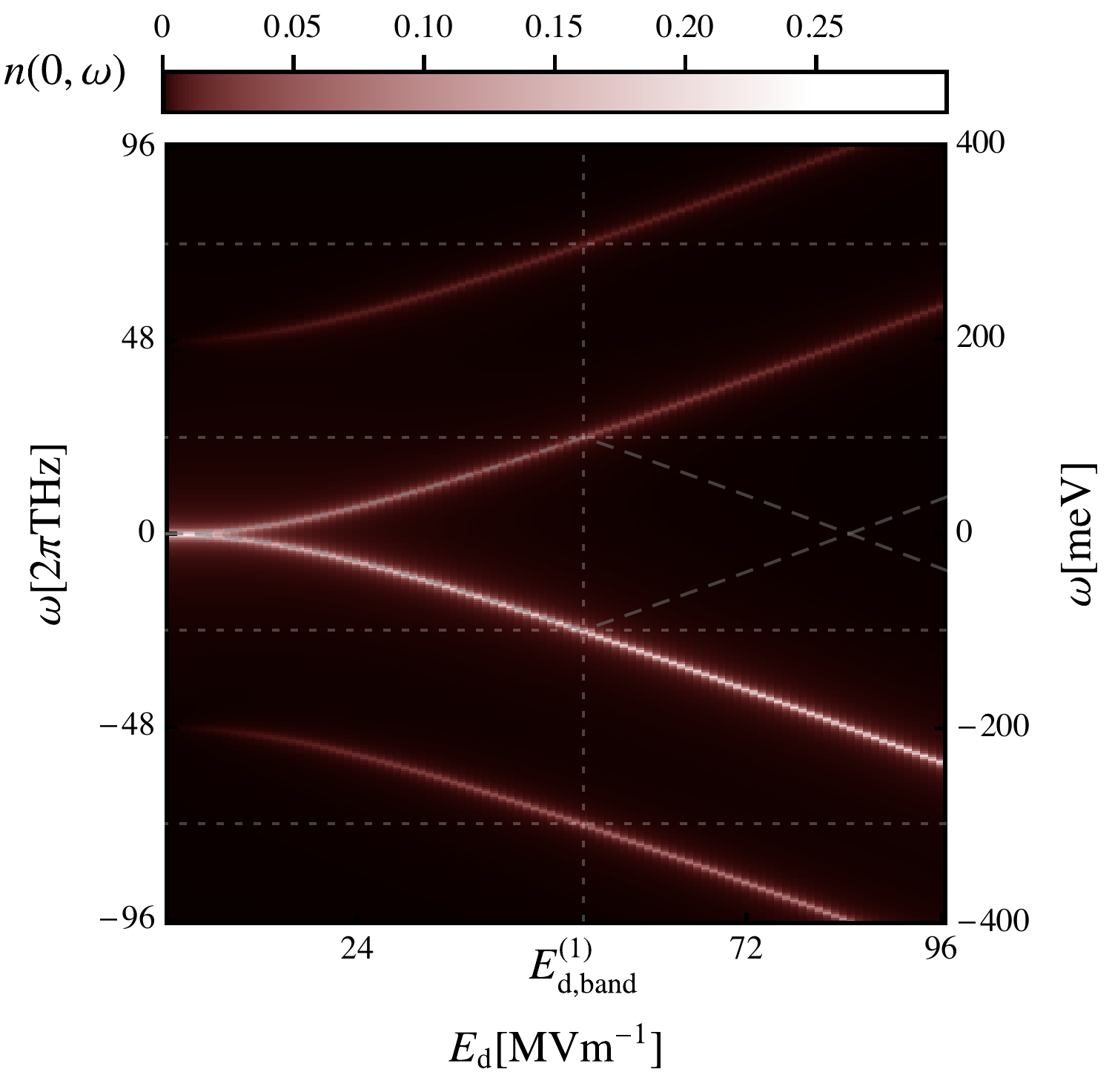}
\caption{
The electron distribution $n(k=0,\omega)$ at the Dirac point as a function of the driving field strength $E_\mathrm{d}$.
The driving frequency is $\omega_\mathrm{d}=2\pi\times 48\si{\tera\hertz}\approx 200\si{\milli\electronvolt}$.
The scaling behavior of the gap at the Dirac point is $\Delta_0= 2\sqrt{v_F^{2}E_\mathrm{d}^2/\omega_\mathrm{d}^2+\omega_\mathrm{d}^2/4}-\omega_\mathrm{d}$.
The vertical dotted line indicates $E_\mathrm{d}=E_{\mathrm{d},\mathrm{band}}^{(1)}.$
The horizontal dotted lines indicate Floquet zone boundaries.
The dashed lines show the Floquet energies at the Dirac point (App.~A) that are formally constrained to be inside the first Floquet zone.
The occupations stay confined within the Floquet bands adiabatically connected to the bare graphene and one replica outwards.
There are no complementary gaps at $k=0$.
}
\label{fig3}
\end{figure}

\begin{figure*}
\centering 
\includegraphics[width=1.0\linewidth]{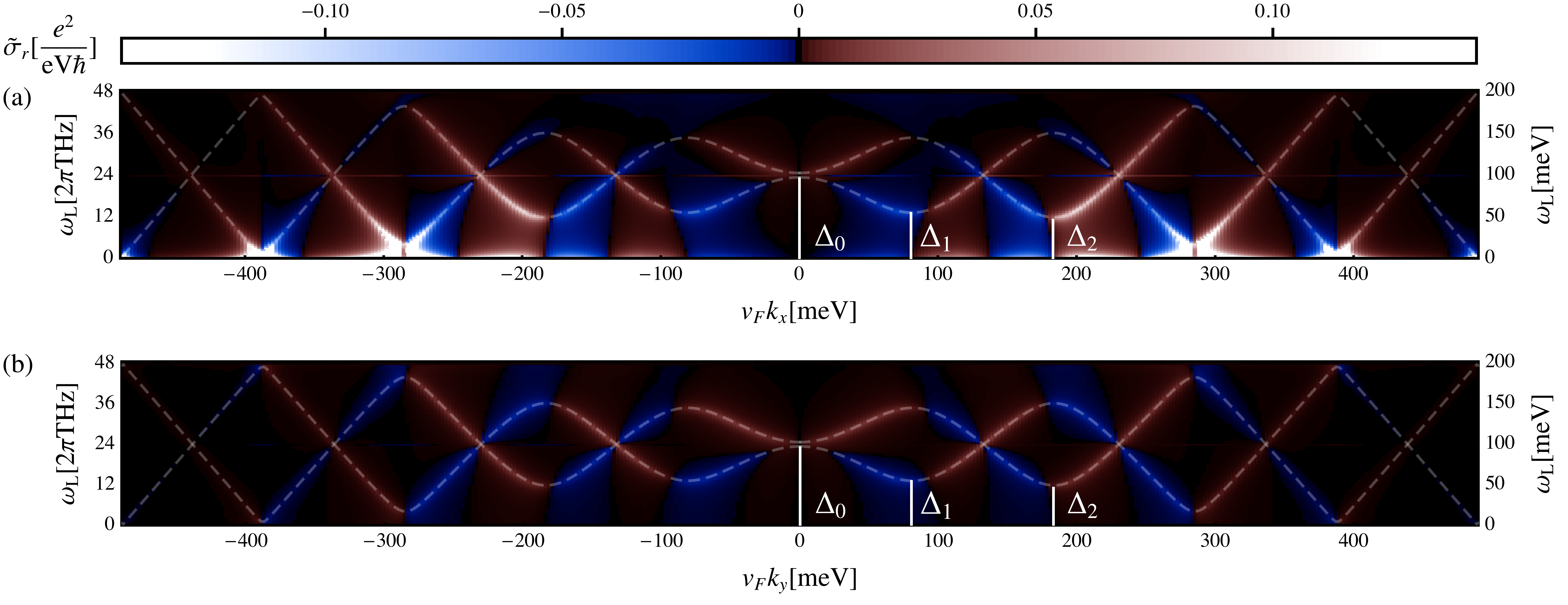}
\caption{The momentum-resolved contributions to the optical conductivity of driven graphene along the $k_x$ (a) and $k_y$ (b) directions.
The driving frequency is  $\omega_\mathrm{d}=2\pi\times 48\si{\tera\hertz}\approx 200\si{\milli\electronvolt}$ and the field strength is $E_\mathrm{d}=34\si{\mega\volt\per\meter}$. 
For these parameters the gap at the Dirac point roughly matches half the driving frequency such that $\Delta_0\approx\omega_\mathrm{d}/2$. The dashed lines indicate the Floquet band energy differences $\Delta\epsilon(k)$ and $\omega_\mathrm{d}-\Delta\epsilon(k)$ (See App.~A).}
\label{fig4}
\end{figure*}

The second observable that we present is the longitudinal optical conductivity.
We propose to measure this quantity experimentally, to compare to the predictions made here. 
In Fig.~\ref{fig2} we show the real part of the optical conductivity as a function of the driving field strength $E_\mathrm{d}$. 
This is obtained from our master equation approach as 
\begin{equation}
\sigma_r(\omega_\mathrm{L}) = \mathrm{Re}{\left[\frac{j_x(\omega_\mathrm{L})}{E_x(\omega_\mathrm{L})}\right]}
\end{equation}
with the longitudinal current and electric field
\begin{align}
j_{x}(\omega_\mathrm{L}) &= n_s n_v ev_F \sum_{\bf k}\int_\tau^{\tau+\frac{2\pi}{\omega_\mathrm{L}}} \mathrm{Tr}(\rho_{\bf k}(t)\sigma_x) e^{i\omega_\mathrm{L} t}\mathrm{d}t\\
E_x(\omega_\mathrm{L}) &= \int_\tau^{\tau+\frac{2\pi}{\omega_\mathrm{L}}} (E_\mathrm{d} \cos(\omega_\mathrm{d} t) + E_\mathrm{L} \sin(\omega_\mathrm{L} t))e^{i\omega_\mathrm{L} t}\mathrm{d}t
\end{align}
where $\tau$ is a point in time where the system has reached its steady state. $n_s=n_v=2$ are the spin- and valley-degeneracies. 
$e$ is the electron charge.
$\sigma_r(\omega_\mathrm{L})$ is obtained for the probing field $E_\mathrm{L}=10\si{\volt\per\meter}$. 
We have verified that the conductivity obtained in this manner is the linear response and that the sum over $k$ includes sufficiently many points surrounding the Dirac point.

As we demonstrate in Fig.~\ref{fig2}, $\sigma_r(\omega_\mathrm{L})$ displays resonant features that match the band gaps of the Floquet spectrum.
The energy gap $\Delta_0$ increases with increasing field strength $E_\mathrm{d}$, in a monotonuous fashion. The energy gaps $\Delta_m$, with $m>0$, first increase with $E_\mathrm{d}$, then reach a maximum at $E_\mathrm{d} = E_{\mathrm{d},\mathrm{max}}^{(m)}$, then decrease, and ultimately reach $0$ at $E_\mathrm{d}=E_{\mathrm{d},\rm{band}}^{(m)}$. 
At this driving strength the gap is located at $k=0$ and merges with $\Delta_0$.

The magnitude of $\sigma_r(\omega_\mathrm{L})$ at the resonance $\Delta_1$, i.e. the magnitude of $\sigma_r(\Delta_1)$, displays a maximum for $E_\mathrm{d} < E_{\mathrm{d},\rm{max}}^{(1)}$, relative to its background, and a minimum for $E_\mathrm{d}>E_{\mathrm{d},\rm{max}}^{(1)}$.
The magnitude of $\sigma_r(\omega_\mathrm{L})$ at $\omega_\mathrm{d} - \Delta_1$, displays the complementary behavior. $\sigma_r(\omega_\mathrm{d} - \Delta_1)$ has a minimum for $E_\mathrm{d} < E_{\mathrm{d},\rm{max}}^{(1)}$, and a maximum for $E_\mathrm{d} > E_{\mathrm{d},\rm{max}}^{(1)}$. 
Note that this does not happen for $\Delta_0$ due to the lack of a complementary gap $\omega_\mathrm{d}-\Delta_0$ as can be seen in Fig.~\ref{fig1} and Fig.~\ref{fig3}.

We obtain analytical expressions for $\Delta_0$ and $E_{\mathrm{d},\mathrm{band}}^{(m)}$ by considering the Hamiltonian in Eq.~\ref{hamiltonian_graphene} at the Dirac point and without probing, i.e. $k=0$ and $E_\mathrm{L}=0$. This has the time-dependent Rabi solutions 
\begin{align}
\ket{+} &\sim e^{i(\omega_\mathrm{d} t/2+\pi/4)} \left(
\begin{array}{cc}
\cos (\Omega t)-\frac{i \omega_\mathrm{d}\sin (\Omega t)}{2\Omega}\\
e^{-i\omega_\mathrm{d} t}\frac{E_\mathrm{d} \sin (\Omega t)}{\Omega \omega}
\end{array}
\right)\\
\ket{-} &\sim e^{-i(\omega_\mathrm{d} t/2+\pi/4)}\left(
\begin{array}{cc}
-e^{i\omega_\mathrm{d} t}\frac{E_\mathrm{d} \sin (\Omega t)}{\Omega \omega_\mathrm{d}} \\
\cos (\Omega t)+\frac{i \omega_\mathrm{d} \sin (\Omega t)}{2\Omega} 
\end{array}
\right)
\end{align}
where 
\begin{equation}
\Omega=\sqrt{\frac{v_F^2 E_\mathrm{d}^2}{\omega_\mathrm{d}^2}+\frac{\omega_\mathrm{d}^2}{4}}. 
\end{equation}
The gap at the Dirac point is given by $\Delta_0 = 2\Omega-\omega_\mathrm{d}$. 
This expression is also the Aharanov-Anandan phase of this system~\cite{Sota}.
In the weak driving limit this gap follows the expected perturbative behavior~\cite{Aoki} 
$
\Delta_0 \approx v_F^2 E_\mathrm{d}^2/\omega_\mathrm{d}^3
$
whereas in the strong driving limit it develops a linear dependence on $E_\mathrm{d}$ as
$
\Delta_0 \approx v_F E_\mathrm{d}/\omega_\mathrm{d}
$.
We use the full expression for $\Delta_0$ to find the driving strengths $E_{\mathrm{d},\rm{band}}^{(m)}$, since they occur whenever the gap $\Delta_0$ spans a multiple of $\omega_\mathrm{d}$. 
By setting $2\Omega-\omega_\mathrm{d}=m\omega_\mathrm{d}$, $m\in\mathbb{N}$, we find
\begin{equation}
E_{\mathrm{d},\rm{band}}^{(m)}~=~v_F^{-1}\sqrt{\frac{m}{2}+\frac{m^2}{4}}\omega_\mathrm{d}^2.
\end{equation}

We display the Dirac gap in Fig.~\ref{fig3}, and compare it to the electron distribution at ${\bf k} = 0$, of the steady state.
We observe that the two maxima of the electron distribution that emerge from $\omega = 0$ follow the prediction of $\pm \Delta_0$, even as $\Delta_0$ grows larger than the Floquet zone boundary at $\omega_\mathrm{d}$. 
Therefore, $\Delta_0$ is a more natural energy scale to predict the resonances at ${\bf k} = 0$ for large driving intensities, than the direct band gap that is strictly smaller than $\omega_\mathrm{d}$.
For increasing field strength $E_\mathrm{d}$, the occupation of the upper two bands decreases. 
The occupation of complementary gaps is zero throughout Fig.~\ref{fig3}.

As visible in Fig.~\ref{fig2}, the conductivity vanishes around the probing frequency $\omega_\mathrm{L}\approx 2\pi\times12\si{\tera\hertz}\approx 50\si{\milli\electronvolt}$ and the driving field strength of $E_\mathrm{d}\approx 39\si{\mega\volt\per\meter}$.
Here the first gap $\Delta_1$ decreases with increasing $E_\mathrm{d}$ and creates a negative contribution that suppresses $\sigma_r(\omega_\mathrm{L})$ to zero. 
For higher order gaps, e.g. $\Delta_2$ and $\Delta_3$ in Fig.~\ref{fig2}, the same phenomenon even leads to a sign change in the conductivity. 
Whenever a gap is in the regime of decreasing with increasing $E_\mathrm{d}$, and no other resonance contributes positively and too strongly to the conductivity, the negative contributions can cancel the background and result in net negative optical conductivity.

In Fig.~\ref{fig4} we resolve the contributions to the conductivity along the $k_x$ and $k_y$ directions relative to the Dirac point in momentum space, defined as 
\begin{equation}
\tilde\sigma_r({\bf k}, \omega_\mathrm{L})=\frac{n_sn_v e v_F|{\bf k}|}{E_x(\omega_{\mathrm{L}})} \int_\tau^{\tau+\frac{2\pi}{\omega_\mathrm{L}}} \mathrm{Tr}(\rho_{\bf k}(t)\sigma_x)e^{i\omega_\mathrm{L} t}  \mathrm{d}t.
\end{equation}
Here we include the linear scaling with the absolute momenta $|{\bf k}|$ in polar coordinates.
Direct interband transitions between neighbouring Floquet bands give rise to resonant features in $\tilde\sigma_r({\bf k}, \omega_\mathrm{L})$ that match the Floquet band energy differences $\Delta\epsilon(k)$ and $\omega_\mathrm{d}-\Delta\epsilon(k)$ (See App.~A).
These resonant features contribute to the conductivity with alternating signs. The sign changes occur close to the band gap locations, but slightly shifted towards (away from) the Dirac point in case the gap size increases (decreases) with respect to the field strength $E_\mathrm{d}$. For gaps that do not change with respect to $E_\mathrm{d}$, i.e. gaps at their maximum, this shift vanishes. Therefore, the accumulated contributions across gaps net either positive or negative conductivity depending on the change in gap size with respect to field strength $E_\mathrm{d}$. 
This is consistent with the enhancements and reductions in $\sigma_r(\omega_\mathrm{L})$ at the gaps $\Delta_m$, with $m>0$, and their complementary gaps $\omega_\mathrm{d}-\Delta_m$, seen in Fig.~\ref{fig2}.

For probing frequencies $\omega_\mathrm{L}$ that are not resonant with a given band gap, $\tilde\sigma_r({\bf k}, \omega_\mathrm{L})$ does not vanish in general. This results in a background conductivity that can obscure the gap $\Delta_0$ at the Dirac point in particular, as is the case for $E_\mathrm{d}<E_{\mathrm{d},\mathrm{max}}^{(1)}$ in Fig.~\ref{fig2}.
Since the band gaps $\Delta_m$, with $m>0$, and their complementary gaps $\omega_\mathrm{d}-\Delta_m$ have a maximum at the field strength $E_\mathrm{d}=E_{\mathrm{d},\mathrm{max}}^{(m)}$, there is always a range that no gap $\Delta_{m}$, with $m>0$, reaches that is centered around $\omega_\mathrm{L}=\omega_\mathrm{d}/2$. 
In this range, it is the gap $\Delta_0$ that is visible predominantly. 
The overall behavior of the gaps is self-similar with respect to the driving frequency $\omega_\mathrm{d}$. Therefore in this system, there always exists a reliable range of probing frequencies where the gap $\Delta_0$ can be observed.

The Floquet interband transitions resonant with $\Delta\epsilon(k)$ occur inside a given Floquet zone.
The ones resonant with $\omega_\mathrm{d}-\Delta\epsilon(k)$ occur across Floquet zone boundaries. Hence, we refer to them as intra-Floquet $\tilde\sigma_r^{\mathrm{intra}}({\bf k},\omega_\mathrm{L})$ and inter-Floquet $\tilde\sigma_r^{\mathrm{inter}}({\bf k},\omega_\mathrm{L})$ contributions to the conductivity, respectively. To distinguish the two we write
\begin{equation} 
\tilde\sigma_r({\bf k},\omega_\mathrm{L}) = 
\tilde\sigma_r^{\mathrm{intra}}({\bf k},\omega_\mathrm{L}) + 
\tilde\sigma_r^{\mathrm{inter}}({\bf k},\omega_\mathrm{L}) + 
\tilde\sigma_r^{\mathrm{bg}}({\bf k},\omega_\mathrm{L}),
\end{equation}
where $\tilde\sigma_r^{\mathrm{bg}}({\bf k},\omega_\mathrm{L})$ is a remaining background contribution accounting for the $\omega_\mathrm{L}\rightarrow 0$ behavior in $\tilde\sigma_r({\bf k},\omega_\mathrm{L})$. 
Fig.~\ref{fig4} (b) shows that $\sigma_r^\mathrm{bg}(k_y,\omega_\mathrm{L})\approx0$.
We fit a function of two Lorentzians located at $\Delta\epsilon(k)$ and $\omega_\mathrm{d}-\Delta\epsilon(k)$ with the same fixed width $\Gamma=1\si{\tera\hertz}$ to the numerical results of $\tilde\sigma_r(k_y,\omega_\mathrm{L})$. 
Specifically, we use
\begin{align}
\tilde\sigma^{\mathrm{fit}}_r(k_y,\omega_\mathrm{L}) &= 
\frac{\Gamma}{\pi} \frac{\tilde\sigma_r^{\mathrm{intra}}(k_y) }{\Gamma^2+(\omega_\mathrm{L}-\Delta\epsilon)^2}\nonumber\\
&+
\frac{\Gamma}{\pi} \frac{\tilde\sigma_r^{\mathrm{inter}}(k_y) }{\Gamma^2+(\omega_\mathrm{L}-\omega_\mathrm{d}+\Delta\epsilon)^2}
\end{align}
as a fitting function.

\begin{figure}[h!]
\centering 
\includegraphics[width=1.0\linewidth]{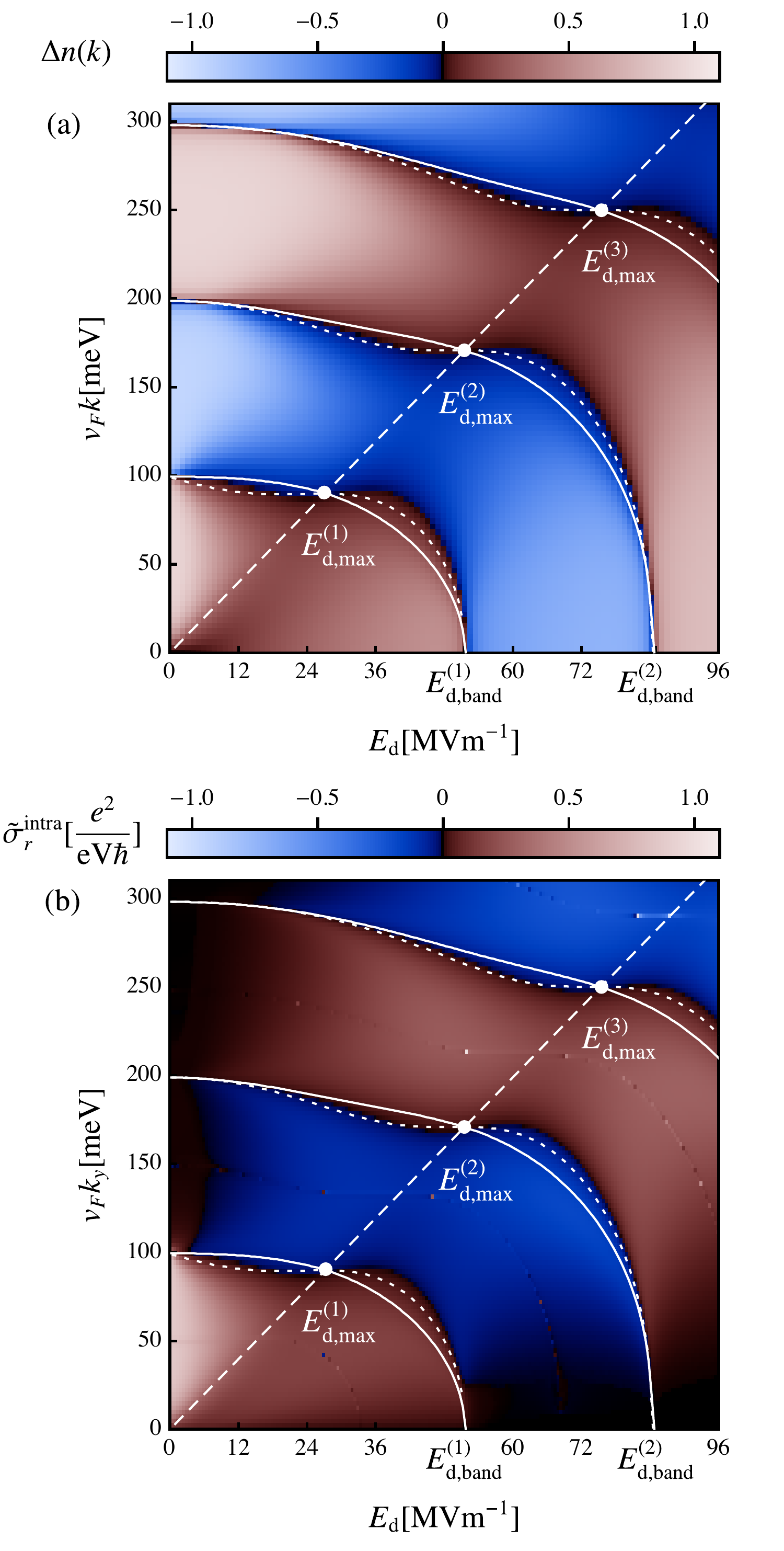}
\caption{The effective occupation $\Delta n(k)$ (a) and the fitted intra-Floquet conductivity $\tilde\sigma_r^{\mathrm{intra}}(k_y)$ (b) as functions of the field strength $E_\mathrm{d}$. 
The solid white lines are given by the locations in momentum space of the band gaps $\Delta_m$, with $m>0$. 
The dotted white lines are given by the zero-crossings of a type of comoving band velocity $\partial_\Pi \Delta \epsilon = 0$ (See App.~A).
The dashed line is given by $k=\frac{E_\mathrm{d}}{\omega_\mathrm{d}}$.}
\label{fig5}
\end{figure}

The conductivity features derive from the transitions between the Floquet bands, and are therefore related to the occupation of these bands.
We define the relative occupation
\begin{equation}
\Delta n(k) = \sum_{m\in\mathbb{Z}} n^-_m(k) - n^+_m(k),
\end{equation}
where $n_m^\pm(k)$ is the occupation at momentum $k$ of the $m$th upper (lower) Floquet band given by integrating $n(k,\omega)$ from $(m-\frac{1}{4}\pm\frac{1}{4})\omega_\mathrm{d}$ to $(m+\frac{1}{4}\pm\frac{1}{4})\omega_\mathrm{d}$. 

Fig.~\ref{fig5} shows the momentum-resolved intra-Floquet conductivity $\tilde\sigma^{\mathrm{intra}}_r(k_y)$ which is determined via fitting as described above, as well as the effective relative occupation $\Delta n(k)$ of the Floquet bands as functions of the field strength $E_\mathrm{d}$. 
They are in good qualitative agreement with each other. 
Both quantities display tongues with alternating signs and zero-crossings seperating them that agree very well between $\Delta n(k)$ and $\tilde\sigma^{\mathrm{intra}}_r(k_y)$. 
The zero-crossings touch the $v_F k$-axis at $m\omega_\mathrm{d}/2$ and the $E_\mathrm{d}$-axis at $E_{\mathrm{d},\mathrm{band}}^{(m)}$, $m\in\mathbb{N}$.
The solid white lines show the location of the Floquet band gaps, i.e. where the radial band velocity vanishes, i.e. $\partial_k\epsilon = 0$. 
They roughly follow the zero-crossings of $\Delta n(k)$ and $\tilde\sigma^{\mathrm{intra}}_r(k_y)$ while showing small, but clear deviations. 
We observe that for $E_{\mathrm{d},\mathrm{max}}^{(m)}<E_\mathrm{d}<E_{\mathrm{d},\mathrm{band}}^{(m)}$ the steady state displays an inversion of the Floquet bands, which creates a negative contribution to the optical conductivity.
The dotted white lines indicate where the comoving radial band velocity $\partial_\Pi\epsilon$ with $\Pi=v_F(k+E_\mathrm{d}/\omega_\mathrm{d})$ vanishes, i.e. $\partial_\Pi\epsilon = 0$.
These lines show improved agreement with the zero-crossings of $\Delta n(k)$ and $\tilde\sigma^{\mathrm{intra}}_r(k_y)$. 
Further, there is an overall resemblance between $\partial_\Pi\epsilon(k)$, $\Delta n(k)$ and $\tilde\sigma^{\mathrm{intra}}_r(k_y)$ (See App.~A).

In conclusion, we have proposed the longitudinal optical conductivity of illuminated graphene as a realistic observable to detect Floquet band gaps. 
We have shown that this quantity displays the Floquet gaps as functions of the driving intensity and the probing frequency. 
In particular, we have pointed out a regime in which the band gap at the Dirac point can be detected. 
All band gaps except for the band gap at the Dirac point, first increase with the driving intensity, approach a maximal value, and then decrease. 
For the increasing regime, the optical conductivity displays a positive contribution. 
For the decreasing regime, it displays a negative contribution that can amount to a total negative conductivity at the given frequency.
We point out that this negative contribution derives from an inversion of the occupation of the Floquet bands. 
Therefore, the proposed experiment not only provides an unambiguous detection of Floquet bands, but also demonstrates dynamical control of transport in solids with light. 

\begin{acknowledgements}
We thank James McIver, Gregor Jotzu and Marlon Nuske for very helpful discussions.
This work is funded by the Deutsche Forschungsgemeinschaft (DFG, German Research Foundation) -- SFB-925 -- project 170620586,
and the Cluster of Excellence 'Advanced Imaging of Matter' (EXC 2056), Project No. 390715994.
\end{acknowledgements}

\bibliography{lit}

\begin{thebibliography}{10}

\bibitem{Hsieh}
D.~N. Basov, R.~D. Averitt, and D.~Hsieh.
\newblock Towards properties on demand in quantum materials.
\newblock {\em Nature Materials}, 16(11):1077--1088, 2017.

\bibitem{Sota}
Takashi Oka and Sota Kitamura.
\newblock Floquet engineering of quantum materials.
\newblock Technical Report~1, 2019.

\bibitem{Schliemann}
Martin Wackerl, Paul Wenk, and John Schliemann.
\newblock Floquet-drude conductivity.
\newblock {\em Phys. Rev. B}, 101:184204, May 2020.

\bibitem{Seradjeh}
Abhishek Kumar, M.~Rodriguez-Vega, T.~Pereg-Barnea, and B.~Seradjeh.
\newblock Linear response theory and optical conductivity of floquet
  topological insulators.
\newblock {\em Phys. Rev. B}, 101:174314, May 2020.

\bibitem{Fiete}
Qi~Chen, Liang Du, and Gregory~A. Fiete.
\newblock Floquet band structure of a semi-dirac system.
\newblock {\em Phys. Rev. B}, 97:035422, Jan 2018.

\bibitem{Demler}
Takuya Kitagawa, Takashi Oka, Arne Brataas, Liang Fu, and Eugene Demler.
\newblock Transport properties of nonequilibrium systems under the application
  of light: Photoinduced quantum hall insulators without landau levels.
\newblock {\em Phys. Rev. B}, 84:235108, Dec 2011.

\bibitem{Galitski}
Netanel~H. Lindner, Gil Refael, and Victor Galitski.
\newblock Floquet topological insulator in semiconductor quantum wells.
\newblock {\em Nature Physics}, 7(6):490--495, 2011.

\bibitem{Rudner}
{Mark S.} Rudner and {Netanel H.} Lindner.
\newblock Band structure engineering and non-equilibrium dynamics in floquet
  topological insulators.
\newblock {\em Nature Reviews Physics}, 2(5):229--244, 5 2020.

\bibitem{Sameti}
Mahdi Sameti and Michael~J. Hartmann.
\newblock Floquet engineering in superconducting circuits: From arbitrary
  spin-spin interactions to the kitaev honeycomb model.
\newblock {\em Phys. Rev. A}, 99:012333, Jan 2019.

\bibitem{Nayak}
Chetan Nayak, Steven~H. Simon, Ady Stern, Michael Freedman, and Sankar
  Das~Sarma.
\newblock Non-abelian anyons and topological quantum computation.
\newblock {\em Rev. Mod. Phys.}, 80:1083--1159, Sep 2008.

\bibitem{Utic}
Igor \ifmmode \check{Z}\else \v{Z}\fi{}uti\ifmmode~\acute{c}\else \'{c}\fi{},
  Jaroslav Fabian, and S.~Das~Sarma.
\newblock Spintronics: Fundamentals and applications.
\newblock {\em Rev. Mod. Phys.}, 76:323--410, Apr 2004.

\bibitem{Haldane}
F.~D.~M. Haldane.
\newblock Model for a quantum hall effect without landau levels:
  Condensed-matter realization of the "parity anomaly".
\newblock {\em Phys. Rev. Lett.}, 61:2015--2018, Oct 1988.

\bibitem{Aoki}
Takashi Oka and Hideo Aoki.
\newblock Photovoltaic hall effect in graphene.
\newblock {\em Phys. Rev. B}, 79:081406, Feb 2009.

\bibitem{Nuske}
M.~Nuske, L.~Broers, B.~Schulte, G.~Jotzu, S.~A. Sato, A.~Cavalleri, A.~Rubio,
  J.~W. McIver, and L.~Mathey.
\newblock Floquet dynamics in light-driven solids.
\newblock {\em Phys. Rev. Research}, 2:043408, Dec 2020.

\bibitem{McIver}
J.~W. McIver, B.~Schulte, F.-U. Stein, T.~Matsuyama, G.~Jotzu, G.~Meier, and
  A.~Cavalleri.
\newblock Light-induced anomalous hall effect in graphene.
\newblock {\em Nature Physics}, 16(1):38--41, November 2019.

\bibitem{Weitenberg}
N.~Fl{\"a}schner, B.~S. Rem, M.~Tarnowski, D.~Vogel, D.-S. L{\"u}hmann,
  K.~Sengstock, and C.~Weitenberg.
\newblock Experimental reconstruction of the berry curvature in a floquet bloch
  band.
\newblock {\em Science}, 352(6289):1091--1094, 2016.

\bibitem{Gedik}
Y.~H. Wang, H.~Steinberg, P.~Jarillo-Herrero, and N.~Gedik.
\newblock Observation of floquet-bloch states on the surface of a topological
  insulator.
\newblock {\em Science}, 342(6157):453--457, 2013.

\bibitem{Pruschke}
J.~K. Freericks, H.~R. Krishnamurthy, and Th. Pruschke.
\newblock Theoretical description of time-resolved photoemission spectroscopy:
  Application to pump-probe experiments.
\newblock {\em Phys. Rev. Lett.}, 102:136401, Mar 2009.

\end{thebibliography}
\bibliographystyle{unsrt}

\newpage
\section*{Appendix A: Floquet energies}

The Floquet energies of a given time-periodic two-level Hamiltonian $H(t)=H(t+\frac{2\pi}{\omega_\mathrm{d}})$ can be obtained by diagonalizing the Hamiltonian 
\begin{equation}
H_F=\begin{pmatrix}
\dots & H_{1} & & & \\
H_{-1}& H_0 + \omega_\mathrm{d} & H_{1}& & \\
& H_{-1}& H_0& H_{1}& \\
& & H_{-1}& H_0 - \omega_\mathrm{d}& H_{1}\\
& & & H_{-1}& \dots\\
\end{pmatrix},
\end{equation}
where $H_n$ are the Fourier components of $H(t)$.
This is done numerically by truncating $H_F$ at a sufficiently large order $n$ of $H_0\pm n\omega_\mathrm{d}$.
The two energies with the smallest absolute values are taken as the upper and lower Floquet energies $\epsilon_+$ and $\epsilon_-$.

\begin{figure}
\centering 
\includegraphics[width=1.0\linewidth]{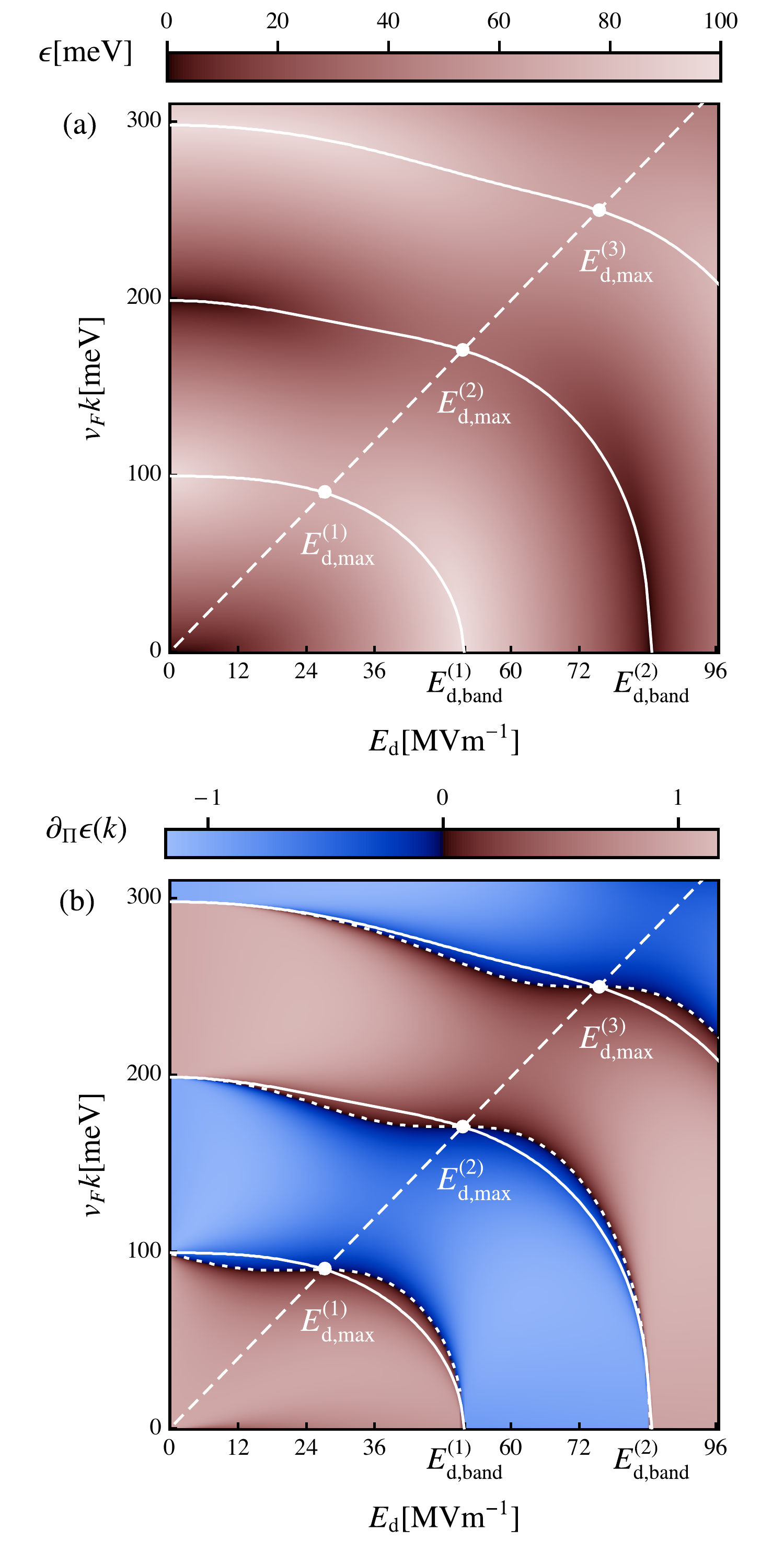}
\caption{
The Floquet band energies $\epsilon$ (a) and comoving band velocity $\partial_\Pi\epsilon$ (b) as functions of the field strength $E_\mathrm{d}$ and for driving frequency $\omega_\mathrm{d}=2\pi\times 48\si{\tera\hertz}\approx 200\si{\milli\electronvolt}$. 
The solid white lines indicate the extrema with respect to momenta which correspond to the locations of the band gaps $\Delta_m$. 
The dashed white lines indicate the zero-crossings of $\partial_\Pi\epsilon$.}
\label{bandprofile}
\end{figure}

For the Hamiltonian Eq.~\ref{hamiltonian_graphene} it is $\epsilon=\epsilon_+=-\epsilon_-$, i.e. $\Delta\epsilon=2\epsilon$.
Fig.~\ref{bandprofile} (a) shows the Floquet band energies $\epsilon$ of this system for $\omega_\mathrm{d}=2\pi\times48\si{\tera\hertz}$ and $E_\mathrm{L} = 0$ as a function of absolute momenta $k$ and driving field strengths $E_\mathrm{d}$.
The band gaps $\Delta_m$, with $m>0$, emerge at momenta equal to multiples of the driving frequency $\omega_\mathrm{d}/2$ and then gradually move in towards the Dirac point where they vanish. 
At these points $E_{\mathrm{d},\rm{band}}^{(m)}$, the respective band gaps $\Delta_m$ merge with the gap $\Delta_0$.
The locations of the gaps $\Delta_m$ are indicated by the white solid lines.
The points $E_{\mathrm{d},\rm{max}}^{(m)}$ indicate the locations of maximal band gaps along those lines.
Note that at the $E_{\mathrm{d},\rm{max}}^{(m)}$ the location of the $m$th gap and the line defined by $k=E_\mathrm{d}/\omega_\mathrm{d}$ intersect.
Fig.~\ref{bandprofile} (b) shows the comoving band velocity $\partial_\Pi \epsilon$, with $\Pi=v_F(k+\frac{E_\mathrm{d}}{\omega_\mathrm{d}})$.
The dashed white lines indicate the zero-crossings of this quantity which deviate from the locations of the band gaps, i.e. the points where $\partial_k\epsilon = 0$.
This comoving band velocity $\partial_\Pi\epsilon(k)$ shows strong similarities to $\Delta n(k)$ and $\tilde\sigma_r^{\mathrm{intra}}(k)$ in Fig.~\ref{fig5}.
\vspace{8cm}
\newpage

\end{document}